\documentclass[twocolumn]{autart}    

\usepackage[pdftex]{graphicx,subcaption}                              
\usepackage{lipsum}
\usepackage[dvips]{epsfig}        
\usepackage{multirow}
\usepackage{epstopdf}
\usepackage{xcolor}
\usepackage{amsmath}
\usepackage{algorithmic}
\usepackage{algorithm}
\usepackage{array}
\usepackage{url}
\usepackage{cite}

\newtheorem{assumption}{Assumption}
\newtheorem{remark}{Remark}
\newtheorem{definition}{Definition}

\begin{document}

\begin{frontmatter}

\title{Zonotope-based Set-membership Parameter Identification of Linear Systems with Additive and Multiplicative Uncertainties and Its Application to Engine Condition Monitoring \thanksref{footnoteinfo}} 

\thanks[footnoteinfo]{Corresponding author H.~Wang. Tel. +1(301)-503-7356.}

\author[HW]{Hao Wang}\ead{autowang@umich.edu},    
\author[IK]{Ilya Kolmanovsky}\ead{ilya@umich.edu},               
\author[HW]{Jing Sun}\ead{jingsun@umich.edu}  

\address[HW]{Department of Navel Architecture and Marine Engineering, University of Michigan, Ann Arbor, MI 48109 USA }  
\address[IK]{Department of Aerospace Engineering, University of Michigan, Ann Arbor, MI 48109 USA}             


\begin{abstract}                          
In this paper, we develop two zonotope-based set-membership estimation algorithms for identification of time-varying parameters in linear models, where both additive and multiplicative uncertainties are treated explicitly. The two recursive algorithms can be differentiated by their ways of processing the data and required computations. The first algorithm, which is referred to as Cone And Zonotope Intersection (CAZI), requires solving linear programming problems at each iteration. The second algorithm, referred to as the Polyhedron And Zonotope Intersection (PAZI), involves linear programming as well as an optimization subject to linear matrix inequalities (LMIs). Both algorithms are capable of providing tight overbounds of the feasible solution set (FSS) in our numerical case studies. Furthermore, PAZI provides an additional opportunity of further analyzing the relation between the estimation results at different iterations. An application to health monitoring of marine engines is considered to demonstrate the utility and effectiveness of the algorithms.
\end{abstract}

\end{frontmatter}

\section{Introduction}
Set-membership estimation has been studied by many researchers since 1990's \cite{Milanese1996}. This approach is often referred to as a guaranteed estimation approach and it generates typically an overbound of the feasible solution set (FSS), which consists of all possible parameters that are consistent with measurements, models, assumptions on noise, and uncertainty bounds. Unlike statistical estimation techniques, no assumptions are made in set-membership estimation about probability distribution of process noise and measurement noise. Recent developments in set-membership estimation include algorithm development and comparison \cite{Cas16}, new insights revealing the connections with statistical estimation \cite{Ble16}, \cite{Wei15}, new techniques for handling nonlinearities \cite{Puig15}, applications to model reference control \cite{Rot14}, \cite{Guo15} and robust MPC \cite{Ping15}, and other novel applications (e.g. simultaneous localization and mapping (SLAM) \cite{Yu15} and diabetes treatment \cite{Her15}). Set-membership techniques for simultaneous input and parameter estimation are developed in \cite{IK06}.

A major topic considered in set-membership estimation is the parameter identification in linear systems. As shown in \cite{Milanese1996}, the FSS of unknown constant parameters can be computed exactly if a linear system with an additive uncertainty is considered. But solving this problem numerically is computationally very involved, thus the approximated feasible solution set (AFSS) is often sought as the over-approximation of the FSS (i.e. FSS $\subseteq$ AFSS). Commonly-used geometric elements for recursively performing such approximations are boxes \cite{Cas16}, ellipsoids \cite{KUR97}, and zonotopes \cite{LE2013}, \cite{J.Bravo2006}. Recently, zonotopes have become popular \cite{LE2013} as the procedures exploiting zonotopes have high computational efficiency and can provide tight overbounds of the FSS. Furthermore, in \cite{J.Bravo2006}, a zonotope-based algorithm is developed for handling the problem of estimating time-varying parameters. This problem is revisited in this paper for the case motivated by with both additive and multiplicative uncertainties while in previous literature (\cite{Milanese1996}, \cite{Cas16}, \cite{LE2013}, and \cite{J.Bravo2006}) only additive uncertainties were treated. The system studied throughout this paper is described as follows.

Consider a linear parametric model in the form treated in \cite{Milanese1996}, \cite{Cas16}, \cite{LE2013}, and \cite{J.Bravo2006}, and generalized to a Multiple-Input-Multiple-Output (MIMO) setting,
\begin{eqnarray}
\label{eqn:0}
y_k = \phi_k^\top \theta_k +u_k,
\end{eqnarray}
where $y_k\in \mathbf{R}^m$ is the known system output, $\theta_k\in \mathbf{R}^n$ represents the true parameter vector to be estimated, $\phi_k \in \mathbf{R}^{n\times m}$ is the regressor and $u_k\in \mathbf{R}^m$ represents the input vector. When $u_k$ in (\ref{eqn:0}) is unknown but has known bounds, we refer to the uncertainty associated with $u_k$ as ``additive'' because of the way it enters the parametric model in (\ref{eqn:0}). There are well-established set-membership estimation algorithms for this case. However, in many cases, such as the engine condition monitoring problem addressed in \cite{HW16} that motivated this study, the uncertainties may affect the regressor, entering the equation (\ref{eqn:0}) in a multiplicative form in relation to the unknown parameters. The literature addressing this case is fairly limited.  In \cite{Cas14}, it is shown that FSS in a problem with constant unknown parameters is, in general, non-convex.  A convex relaxation approach can be pursued with box-type solutions for FSS, which may lead to very conservative over-approximations in some cases.  A convex relaxation approach to a similar problem is pursued in \cite{Piga12}, where the problem involving multiplicative uncertainties is referred as an Error-in-Variable (EIV) problem. The setting in \cite{Piga12} is different from ours in that \cite{Piga12} explicitly handles the dependence of the regressor on past outputs while in our health monitoring applications this dependence does not appear, and hence we focus on the latter case. When applied in a setting of \cite{Piga12}, our algorithms may provide more conservative results as they do not use extra modeling information.

In this paper, we treat the set-membership identification problem of time-varying parameters in linear models and we account for both additive uncertainties in $u_k$ and multiplicative uncertainties in $\phi_k$. Under our assumptions, including boundedness of the time-varying parameters, the FSS is convex and can be computed by a recursive formula, which is formed by intersecting a prior estimate and a polyhedron. This polyhedron is defined by an information set, which consists of all the constraints on the feasible parameters obtained from the $m$ measurements at the current time step. 

In order to build corresponding AFSS for the FSS at each time step, two algorithms are developed in this paper. They are distinguished by their ways of processing the $m$ measurements in $y_k$. These $m$ measurements are segmented into $l$ subsets ($l\leq m$), each of which is referred to as a mini-batch in this paper. Within one iteration, the first algorithm processes a single measurement in $y_k$, which defines a convex cone constraint on the feasible parameters. Then, this algorithm computes candidate zonotopes that overbound the intersection between the prior estimate and the cone. Among all those candidates, the one with minimal estimated volume is selected and used as a prior estimate for the next iteration. After $m$ iterations, all the measurements are processed and an AFSS for the FSS at the current time step is obtained. For handling time-varying parameters, the updated AFSS is propagated forward based on the rate of variations, providing a prior estimate of the AFSS for the next time step. This algorithm is referred as CAZI which stands for Cone And Zonotope Intersection. In contrast to the first algorithm, the second algorithm processes multiple measurements in each iteration. This algorithm focuses on building a zonotope that overbounds the intersection between the prior estimate and a polyhedron, which is defined by the constraints associated with the current mini-batch of measurements. The number of measurements collected in each mini-batch is limited in order to reduce the computational complexity. An optimization problem subject to constraints prescribed by linear matrix inequalities (LMIs) is subsequently derived for computing the AFSS at each iteration. The P-radius \cite{LE2013} of the resulting zonotope is minimized by solving this optimization problem. The propagations of the solutions from one iteration to another are as in the CAZI algorithm. Since the second algorithm emphasizes the way of building overbounds on Polyhedron And Zonotope Intersection, we refer it as PAZI algorithm. Furthermore, PAZI algorithm is amenable to the analysis of the AFSS evolution over iterations. It is also found that the feasibility of the LMIs is closely related to signal richness and uncertainty level of the measurements. For illustrating the algorithms, an application to the engine condition monitoring problem with unknown health parameters is considered.

Our earlier conference paper \cite{HW17} has attempted to solve the identification problem with single measurement at each time step (i.e. $y_k\in \mathbf{R}$) and has focused on the CAZI algorithm. In this paper, this problem is generalized to the multiple-measurement case, a new PAZI algorithm is introduced to handle this problem and the estimation performance is compared with the one of the CAZI algorithm.

The rest of the paper is organized as follows. Mathematical preliminaries are reviewed in Section~\ref{sec:1}. The problem formulation and the FSS properties are discussed in Section~\ref{sec:2}. The CAZI and PAZI algorithms are introduced in Section~\ref{sec:3} and Section~\ref{sec:4}, respectively. In Section~\ref{sec:5}, the two algorithms are applied to an engine condition monitoring problem. The conclusions are presented in Section~\ref{sec:6}.


\section{Preliminaries}\label{sec:1}
Before we proceed with the detailed problem formulation, the following definitions are given since they are used throughout the paper.
\begin{definition}{(Polyhedron)}:
	A polyhedron is a convex set defined by intersecting a finite number of half spaces,  $\mathcal{P}=\{\hat{\theta}\in \mathbf{R}^n: A\hat{\theta}\leq b\}$, where $A\in\mathbf{R}^{m\times n}, b\in \mathbf{R}^m$. 
\end{definition}
\begin{definition}{(Polytope)}:
	A polytope is a bounded polyhedron.
\end{definition}
\begin{definition}{(Minkowski sum)}:
	The Minkowski sum of two sets $X$ and $Y$, denoted as $X\oplus Y$, is a set defined as $X\oplus Y=\{z:\exists x\in X, y\in Y~\text{such that}~ z=x+y\}$.
\end{definition}
\begin{definition}{(Unit hypercube of order $m$)}:
	A unit hypercube of order $m$, $\mathbf{B}^m$, is a set of $m$-dimensional vectors defined by $\mathbf{B}^m=\{b\in \mathbf{R}^m: ||b||_\infty\leq 1\}$.
\end{definition}
\begin{definition}{(Zonotope of order $m$)}:
	Given a vector $p\in \mathbf{R}^n$ and a matrix $H\in \mathbf{R}^{n\times m}$, a zonotope of order $m$ is a set of $n$-dimensional vectors defined by \cite{LE2013}
	\begin{eqnarray}
	\label{zono:def1}
	\mathcal{Z}=\{\hat{\theta}\in \mathbf{R}^n: \hat{\theta}\in p\oplus H\mathbf{B}^m\},
	\end{eqnarray} where $H\mathbf{B}^m=\{Hz: z\in \mathbf{B}^m\}$ is a linear projection of $\mathbf{B}^m$ into $n$-dimensional parameter space.
\end{definition} 
The approach of representing a zonotope by (\ref{zono:def1}) is often referred to as affine projection (transformation) of a hypercube. Geometrically, zonotopes are a special class of polytopes, thus (\ref{zono:def1}) can be transformed into an equivalent half-space representation as follows,
\begin{eqnarray}
\mathcal{Z}=\{\hat{\theta}\in \mathbf{R}^n: A\hat{\theta}\leq b\}.\label{zono:def2}
\end{eqnarray} 
 Since their different formats can be exploited in different ways, both representations will be used at different algorithm development stages: (\ref{zono:def1}) is used to implement set operations while (\ref{zono:def2}) is used for optimization. 
\begin{definition}{(P-radius of a zonotope)}:
	The P-radius of a zonotope $\mathcal{Z}=p\oplus H\mathbf{B}^m\subset \mathbf{R}^n$ is given by the following expression \cite{LE2013}:
	\begin{eqnarray}
	r&=&\max_{z\in\mathcal{Z}}(||z-p||^2_P)=\max_{z \in \mathbf{B}^m}~\|Hz\|_P^2 \nonumber \\
	&=&\max_{z\in \mathbf{B}^m}(z^\top H^\top P Hz),\label{eqn:def2}
	\end{eqnarray}
	where $P$ is an n-dimensional positive definite matrix.
\end{definition}  
\begin{definition}{(Strip)}:
	Given vectors $c\in\mathbf{R}^n$, $d\in\mathbf{R}$, and $\sigma\in \mathbf{R}$, a strip set $\mathcal{S}$ is defined by,
	\begin{eqnarray}
	\mathcal{S}=\{\hat{\theta}\in \mathbf{R}^n: |c^\top\hat{\theta}-d|\leq \sigma\},\label{strip:def}
	\end{eqnarray} 	
\end{definition}
where $c$ and $\sigma$ are referred to as the orientation and the size of the strip. 

\section{Problem formulation}\label{sec:2}
The following set-membership identification problem of time-varying parameters is considered in this paper:

For a discrete-time linear parametric model (\ref{eqn:0}) with both additive and multiplicative uncertainties in $u_k$ and $\phi_k$, and time-varying parameters $\theta_k$ satisfying $|\theta_k-\theta_{k-1}|\leq \gamma_k$, given $u^l_k$, $u^u_k$， $\phi^l_k$, $\phi^u_k$, $y_k$ and $\gamma_k$, where $k\in K=\{1,2,3....N\}$ is the index for the data point, find the largest set for each $k$, denoted as $\mathcal{C}_k\subset\mathbf{R}^n$, such that for all $\hat{\theta}\in \mathcal{C}_k$, there exist $\hat{\phi}_k\in \mathbf{R}^{n\times m}$ and $\hat{u}_k\in \mathbf{R}^m$ which satisfy
\begin{eqnarray}
\label{eqn:1}
&y_k& = \hat{\phi}_k^\top \hat{\theta} + \hat{u}_k,\nonumber\\
\label{eqn:2}
&\phi^l_k&\leq \hat{\phi}_k \leq \phi^u_k,\nonumber\\
\label{eqn:3}
&u^l_k&\leq \hat{u}_k \leq u^u_k,\nonumber \\
&\hat{\theta}&\in \hat{\mathcal{C}}_{k}=\mathcal{C}_{k-1}\oplus \Gamma_k\mathbf{B}^n,
\end{eqnarray}
where\footnote{The inequalities in (\ref{eqn:1}) are element-by-element.} $u^l_k$ and $u^u_k$ characterize the additive uncertainty, $\phi^l_k$ and $\phi^u_k$ characterize the multiplicative uncertainty, $y_k$ represents the measurement, $\hat{\theta}$, $\hat{\phi}_k$, and $\hat{u}_k$ represent feasible estimates of the true values of the unknown parameters, regressors and input, respectively. $\hat{\mathcal{C}}_{k}$ in the last constraint of (\ref{eqn:2}) is the ``predicted'' estimated parameter set, which is obtained based on the result from previous step $\mathcal{C}_{k-1}$ ($\mathcal{C}_{0}$ is the initial guess based on a priori information) and a priori known bounds on parameter rate of change. In (\ref{eqn:1}), $\oplus$ represents the Minkowski sum, $\Gamma_k$ is an $n$-dimensional diagonal matrix given by $\Gamma_k=diag(\gamma_k)$ and  $\mathbf{B}^n$ is an $n$-dimensional unit hypercube. Note that, any $\hat{\theta}=[\hat{\theta}_{1},\hat{\theta}_{2},...\hat{\theta}_{n}]^\top\in \mathcal{C}_k$ represents a feasible estimate of the true parameter $\theta_k=[\theta_{1,k},\theta_{2,k},...\theta_{n,k}]^\top$. 

According to \cite{Milanese1996}, the set $\mathcal{C}_k$ defined above represents the FSS of our problem at each time step, denoted as FSS$_k$. For describing how to recursively compute the FSS$_k$, the information set $\mathcal{L}_k$ is defined as follows.

\begin{assumption}
	Each component of the true parameter vector $\theta_k$ is non-negative, $\theta_k\geq 0$, for all $k\in \{1,2,3....N\}$.
\end{assumption}

\begin{definition}{(Information set)}: The information set $\mathcal{L}_k$ is a set of all feasible parameters, that are consistent with the model (\ref{eqn:0}), the measurements $y_k$, and the uncertainty bounds at time step $k$, namely:
	\begin{eqnarray}
	\mathcal{L}_k=\{\hat{\theta}: \phi^{u\top}_k\hat{\theta} \geq y_k-u_k^u,~\phi^{l\top}_k\hat{\theta}\leq y_k-u_k^l,\hat{\theta}\geq 0\}.\label{eqn:5}
	\end{eqnarray}
\end{definition}
Geometrically, $\mathcal{L}_k$ represents a polyhedron in the parameter space.
\begin{remark}
	Assumption 1 is not restrictive. When this assumption is not satisfied, the problem can be easily reformulated by a simple variable transformation. For example, suppose there exist bounds $d_i$, $i=1,\cdots, n$, such that
	\begin{eqnarray}
	d_i\geq\max_{k}(|\theta_{i,k}|), i=1,2,...,n.\label{eqn:8}
	\end{eqnarray}
	Then the information set can be re-defined as
	\begin{eqnarray}
	\mathcal{L}_k=\{\bar{\theta}: \phi^{u\top}_k\bar{\theta}\geq y_k-\bar{u}_k^u,~\phi^{l\top}_k\bar{\theta}\leq y_k-\bar{u}_k^l,\bar{\theta}\geq 0\},\label{eqn:6}
	\end{eqnarray}
	where \begin{eqnarray}
	\bar{\theta}&=&\hat{\theta}+d,\nonumber\\
	\bar{u}_k^u&=&u_k^u-d^\top\phi^l_k,\nonumber\\
	\bar{u}_k^l&=&u_k^l-d^\top\phi^u_k.\label{eqn:7}
	\end{eqnarray}
\end{remark} 
The problem can now be reformulated so that instead of estimating $\theta_k$, we are estimating the non-negative parameter $\bar{\theta}$, which satisfies Assumption 1.

Following the approach in \cite{J.Bravo2006}, the convex solution set, FSS$_k$, can be obtained recursively as 
\begin{eqnarray}
\text{FSS}_k=\text{FSS}_k^-\cap \mathcal{L}_k=(\text{FSS}_{k-1}\oplus \Gamma_k\mathbf{B}^n)\cap \mathcal{L}_k,\label{eqn:9}
\end{eqnarray} where FSS$_k^-$ represents a prior estimate of FSS$_k$.
\begin{remark}
	If unknown parameters are constant, then $\gamma_k=0$, FSS$_{k-1}$$\supseteq$FSS$_{k}$ for all $k\in K$, and (\ref{eqn:9}) can be simplified as
	\begin{eqnarray}
	\text{FSS}_k=\text{FSS}_{k-1}\cap \mathcal{L}_k.\label{eqn:10}
	\end{eqnarray} 
\end{remark}
In general, computing FSS$_k$ is a difficult task. Hence AFSS$_k$, the over approximation of the corresponding FSS$_k$, is subsequently exploited.  In particular, we use zonotopes as they can provide tight overbounds and lead them to efficient manipulations \cite{LE2013}. The basic approach exploits the following set inclusion,
\begin{eqnarray}
\text{AFSS}_k \supseteq \text{AFSS}_k^-\cap \mathcal{L}_k'=(\text{AFSS}_{k-1}\oplus \Gamma_k\mathbf{B}^n)\cap \mathcal{L}_k',\label{eqn:9compli}
\end{eqnarray}
where AFSS$_k$, AFSS$_k^-$, AFSS$_{k-1}$ and $\mathcal{L}_k'$ represent the supersets of FSS$_k$, FSS$_k^-$, FSS$_{k-1}$ and $\mathcal{L}_k$ in (\ref{eqn:9}), respectively. In our proposed algorithm, the sets AFSS$_k$, AFSS$_k^-$ and AFSS$_{k-1}$ are zonotopes of limited complexity such that the computations can be simplified. We define $\mathcal{L}_k'$, with $\mathcal{L}_k \subseteq \mathcal{L}_k'$, as follows,   
\begin{eqnarray}
\mathcal{L}_k'=\{\hat{\theta}: {\phi^{u\top}_k}\hat{\theta}\geq y_k-u_k^u, {\phi^{l\top}_k}\hat{\theta}\leq y_k-u_k^l\}.\label{eqn5:compli}
\end{eqnarray}

However, computing AFSS$_k$ directly from (\ref{eqn:9compli}) can still be challenging considering the complexity introduced by multiple measurements in $y_k$. Thus two algorithms for handling the multiple measurements are proposed in the following sections. The first one (CAZI) processes single measurement in each iteration while the second one (PAZI) processes multiple measurements as a mini-batch.

\section{Cone And Zonotope Intersection (CAZI) algorithm}\label{sec:3}
CAZI algorithm splits the task of computing AFSS$_k$ in (\ref{eqn:9compli}) into $m$ subtasks, each of which can be considered as computing an intermediate estimate, and is represented as follows,
\begin{eqnarray}
\text{AFSS}_{k,i} \supseteq \text{AFSS}_{k,i-1}\cap \mathcal{L}_{k,i}',\label{eqn:15}
\end{eqnarray}
where $i=1,2,...,m$ is the index of the intermediate estimate at time step $k$, $\text{AFSS}_{k,0}=\text{AFSS}_{k-1,m}\oplus \Gamma_k\mathbf{B}^n$ represents a prior estimate at the current time step, $\text{AFSS}_{k,m}=\text{AFSS}_{k}$ represents a posterior estimate at the current time step, $\mathcal{L}'_{k,i}$ is the information set defined by a single measurement as
\begin{eqnarray}
\mathcal{L}_{k,i}'=\{\hat{\theta}: \phi^{u\top}_{k,i}\hat{\theta}\geq y_{k,i}-u_{k,i}^u,\nonumber\phi^{l\top}_{k,i}\hat{\theta}\leq y_{k,i}-u_{k,i}^l\},\label{eqn5:compli_2}
\end{eqnarray}
where $\phi^u_{k,i}$, $\phi^l_{k,i}$ are the $i$th columns of $\phi^u_{k}$, $\phi^l_{k}$, respectively, and $y_{k,i}$, $u^u_{k,i}$, and $u^l_{k,i}$ are the $i$th entries of $y_{k}$, $u^u_{k}$, and $u^l_{k}$, respectively.

As defined in \cite{LB69}, if the following set corresponding to each $\mathcal{L}'_{k,i}$, $\mathcal{V}=\{\hat{\theta}: \phi^{u\top}_{k,i}\hat{\theta}= y_{k,i}-u_{k,i}^u,~\phi^{l\top}_{k,i}\hat{\theta}= y_{k,i}-u_{k,i}^l\}$, is not empty, then $\mathcal{L}'_{k,i}$ is a convex cone with vertex in $\mathcal{V}$.

The detailed algorithm can be summarized as \textbf{Algorithm 1}:
\begin{algorithm}[h]
	\caption{CAZI Algorithm \label{alg:cazi}}
	\begin{algorithmic}[1]
		\STATE{Set $k=1$;}
		\WHILE{$k\leq N$}
		\STATE{set $i=1$;}
		\STATE{collect measurement vector $y_k$, update signal bounds $u^l_k$, $u^u_k$， $\phi^l_k$, $\phi^u_k$;} %
		\WHILE{$i\leq m$}%
		\STATE{formulate the set $\mathcal{L}_{k,i}'$;}%
		\STATE{build support strips, $\mathcal{S}_{k,i}^1$ and $\mathcal{S}_{k,i}^2$, that overbound AFSS$_{k,i-1}$ $\cap$ $\mathcal{L}_{k,i}'$;}
		\STATE{obtain candidate zonotopes that overbound AFSS$_{k,i-1}$ $\cap$ $\mathcal{S}_{k,i}^1$ and AFSS$_{k,i-1}$ $\cap$ $\mathcal{S}_{k,i}^2$, respectively;}
		\STATE{among all candidates, set AFSS$_{k,i}$ as the one with minimal estimated volume, which is denoted as $\mathcal{Z}_{k,i}^*=p_{k,i}^*\oplus H_{k,i}^*\mathbf{B}^r$;}
		\STATE {set i=i+1;}
		\ENDWHILE %
		\STATE {set AFSS$_{k+1,0}$=$\mathcal{Z}_{k,m}^*\oplus\Gamma_{k+1}\mathbf{B}^n=p_{k,m}^*\oplus [H_{k,m}^*~\Gamma_{k+1}]\mathbf{B}^{r+n}$;
		\STATE{set $k=k+1$;}}
		\ENDWHILE
	\end{algorithmic}
\end{algorithm}

The procedure for bounding the intersection between  AFSS$_{k,i-1}$ and $\mathcal{L}_{k,i}$ is described below.
\subsection{Building support strips} 
\begin{definition}{(Support strip)}:
	Given a polytope and a vector $c\in \mathbf{R}^n$ (i.e., $c$ defined in (\ref{strip:def})), the support strip is the strip with orientation $c$ and minimal size, which contains this polytope.
\end{definition}

In our algorithm, we are exploiting support strips that overbound AFSS$_{k,i-1}$ $\cap$ $\mathcal{L}_{k,i}'$, which is a polytope. In particular, the orientations of the two support strips are selected based on the two hyperplanes forming $\mathcal{L}_{k,i}'$. The sizes of the two strips are minimized according to the following proposition.  

\textbf{Proposition 1}: Given a polytope $\mathcal{P}=\{\hat{\theta}: A\hat{\theta}\leq b, A\in\mathbf{R}^{m\times n}, b\in \mathbf{R}^m\}$ and a convex set $\mathcal{C}=\{\hat{\theta}: \phi^\top_1\hat{\theta}\geq b_1, \phi^\top_2\hat{\theta}\leq b_2, \hat{\theta}\in \mathbf{R}^n, \phi_1\neq 0\in \mathbf{R}^n, \phi_2 \neq 0\in \mathbf{R}^n\}$. If $\mathcal{P}\cap \mathcal{C} \neq \emptyset$,  then the following computations give the smallest value of $\delta=\delta_1^*$ for constructing the strip $\mathcal{S}^1$ such that $\mathcal{P}\cap \mathcal{C} \subseteq \mathcal{S}^1$, where
\begin{eqnarray}
\label{eqn:14}
\mathcal{S}^1=\{\hat{\theta}: \left|\phi_1^\top\hat{\theta}-(b_1+{\delta\over 2}\|\phi_1\|)\right|\leq {\delta\over 2}\|\phi_1\|\}.
\end{eqnarray}
The value of $\delta_1^*$ is obtained by
\[
\tag{P1}
\begin{array}{rrcl}
\delta_1^*&=&\max\limits_{\hat{\theta}}|  {\tfrac{(\hat{\theta}-\tilde{\theta}_1)^\top\phi_1}{\|\phi_1\|}}|           \\
&\text{s.t.} &  \hat{\theta}     \in  \mathcal{P}\cap \mathcal{C}, 
\end{array}
\]
where $\phi^\top_1\tilde{\theta}_1= b_1$. Similarly, $\mathcal{S}^2$ is given by (\ref{eqn:14}) with $\phi_2$ replacing $\phi_1$, $b_2$ replacing $b_1$, and $\delta=\delta_2^*$ defined by (P1) with $\phi_1$ replaced by $\phi_2$ and $\tilde{\theta}_2$ replacing $\tilde{\theta}_1$ where $\phi^\top_2\tilde{\theta}_2= b_2$. 

Detailed proof of this proposition will be shown in the Appendix.
\begin{figure}[h!]
	\begin{center}
		\includegraphics[width=7cm]{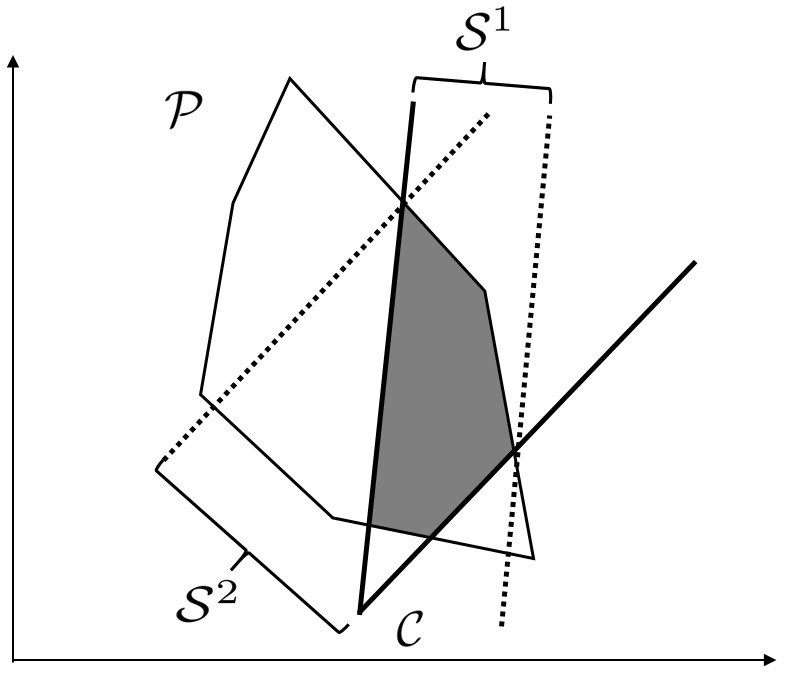}    
		\caption{Example of support strips, $\mathcal{S}^1$ and $\mathcal{S}^2$, which overbound $\mathcal{P}$ $\cap$ $\mathcal{C}$ shown in grey.} 
		\label{fig:illustration_supportstrip}
	\end{center}
\end{figure} 

Applying this proposition, if we set $\mathcal{P}$=AFSS$_{k,i-1}$ and $\mathcal{C}=\mathcal{L}_{k,i}'$, two support strips can be built based on the two hyperplanes that define $\mathcal{L}_{k,i}'$. The corresponding $\delta_1^*$ and $\delta_2^*$ are determined by solving two linear programming problems of form (P1). Fig.~\ref{fig:illustration_supportstrip} illustrates the procedure of building the two support strips.

\subsection{Building candidate zonotopes and the selection of $\mathcal{Z}_k^*$}
Given the two support strips, candidate zonotopes which overbound the intersection between a zonotope and each support strip can be constructed using parameterization from \cite{J.Bravo2006}. Each of these zonotopes, therefore, also overbounds AFSS$_{k,i-1}$ $\cap$ $\mathcal{L}_{k,i}'$. The parametrization of a zonotope overbounding an intersection of a zonotope and a strip is as follows.

Consider a zonotope $\mathcal{Z}=p\oplus H\mathbf{B}^r=p\oplus [H^1 H^2...H^r]\mathbf{B}^r\subset \mathbf{R}^n$ and a strip $\mathcal{S}=\{\hat{\theta}\subset \mathbf{R}^n: |c^\top\hat{\theta}-d|\leq \sigma\}$. Then \cite{J.Bravo2006} for every integer $j$, $0\leq j\leq r$,
\begin{eqnarray}
\label{eqn:18}
\mathcal{Z}\cap \mathcal{S} \subseteq \mathcal{Z}_j=v(j)\oplus T(j)\mathbf{B}^r,
\end{eqnarray}
where 
\begin{eqnarray}
\label{eqn:19}
v(j)=\begin{cases}
p+(\frac{d-c^\top p}{c^\top H^j})H^j, \text{if $1\leq j\leq r$ and $c^\top H^j\neq 0$} \\ 
p, ~~~~~~~~~~~~~~~~~~~~~~~~~~~~~~~~~~~~~\text{otherwise}
\end{cases},\nonumber
\end{eqnarray}
\begin{eqnarray}
\label{eqn:20}
T(j)&=&\begin{cases}
[T^j_1T^j_2...T^j_r], \text{if $1\leq j\leq r$ and  $c^\top H^j\neq 0$} \\ 
H, ~~~~~~~~~~~~~~~~~~~~~~~~~~~~~~~\text{otherwise}
\end{cases},\nonumber\\
T_i^j&=&\begin{cases}
H^i-(\frac{c^\top H^i}{c^\top H^j})H^j, \text{if $i\neq j$} \\ 
(\frac{\sigma}{c^\top H^j})H^j,\text{if $i=j$}
\end{cases}.
\end{eqnarray}

\begin{figure}[h!]
	\begin{center}
		\includegraphics[width=8cm]{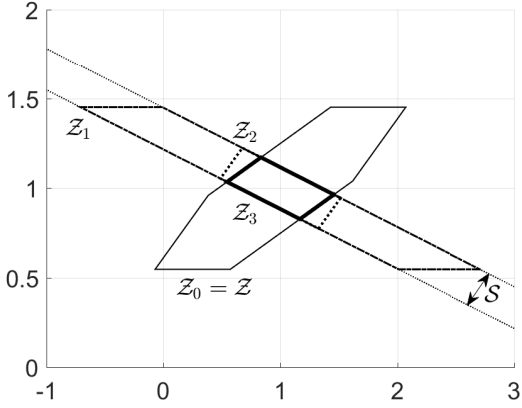}    
		\caption{Example of candidate zonotopes (from $\mathcal{Z}_0$ to $\mathcal{Z}_3$).} 
		\label{fig:build_zonos}
	\end{center}
\end{figure}

As an example, consider a zonotope and a strip defined as follows,
\begin{eqnarray}
\mathcal{Z}=\left[\begin{array}{l} \displaystyle
1 \\ 
3
\end{array}\right]\oplus \left[\begin{array}{lll} \displaystyle
0.3212 & 0.2268 & 0.5235 \\ 
0 & 0.2063 & 0.2467
\end{array} \right]\mathbf{B}^3, \nonumber
\end{eqnarray}
\begin{eqnarray}\displaystyle
\label{eqn:strip}
\mathcal{S}=\{\hat{\theta}: |[1~3]~\hat{\theta}-4|\leq 0.35\}.
\end{eqnarray}
Fig.~\ref{fig:build_zonos} gives an example of the candidate zonotopes that overbound $\mathcal{Z}\cap \mathcal{S}$.   

In CAZI algorithm, $\mathcal{S}$ is set to be $\mathcal{S}_{k,i}^1$ and  $\mathcal{S}_{k,i}^2$ that are built at Step 7. Then, there are, in total, $2(r_{k,i-1}+1)$ candidate zonotopes built for each iteration based on the parametrization described above, where $r_{k,i-1}$ is the order of the zonotope AFSS$_{k,i-1}$. Among all these candidates, $\mathcal{Z}_{k,i}^*$ is chosen to be the one with the smallest estimated volume. The volume of a zonotope can be easily estimated, see \cite{J.Bravo2006} for details. 

\subsection{Reducing the order of AFSS$_{k+1,0}$}
Step 12 of our proposed CAZI algorithm, i.e., formulating AFSS$_{k+1,0}$ based on $Z^*_{k,m}$, will cause the order of the resulting zonotope to increase when identifying time-varying parameters. To avoid the increase in zonotope complexity, we limit the order of  AFSS$_{k+1,0}$ which is an over-approximation for FSS$_{k+1}$ to be $n$. Note that $n$ is the lowest possible order of a non-zero volume zonotope in an $n$-dimensional parameter space. See \cite{LE2013} for different approaches to reducing the order of a zonotope. They involve sorting the columns of $H$ matrix of a zonotope and simple algebraic calculations.  

\section{Polyhedron And Zonotope Intersection (PAZI) algorithm}\label{sec:4}
Compared with CAZI, PAZI algorithm splits the task of computing AFSS$_k$ in (\ref{eqn:9compli}) into $l$ subtasks, where $l$ is the number of mini-batches determined at each time step. The $m$ measurements are distributed into these $l$ mini-batches, and each mini-batch may consist of more than one measurement. The intermediate estimates are computed recursively as: 
\begin{eqnarray}
\text{AFSS}_{k,i} \supseteq \text{AFSS}_{k,i-1}\cap \mathcal{P}_{k,i},\label{eqn:24_new}
\end{eqnarray}
where $i=1,2,...,l$ is the index of the mini-batch at time step $k$, $\text{AFSS}_{k,l}=\text{AFSS}_{k}$ represents the estimate at the current time step $k$, $\mathcal{P}_{k,i}=\bigcap_j\mathcal{L}_{k,j}', j=1,2,...,M_i$ is the information set defined by measurements included in the current mini-batch $i$, where $j$ and $M_i$ are the index and the number of measurements in the current mini-batch $i$, respectively.

The detailed PAZI algorithm is summarized as \textbf{Algorithm 2}:
\begin{algorithm}[h]
	\caption{PAZI Algorithm \label{alg:pazi}}
	\begin{algorithmic}[1]
		\STATE{Set $k=1$;}
		\WHILE{$k\leq N$}
		\STATE{set $i=1$;}
		\STATE{collect measurement vector $y_k$, update signal bounds $u^l_k$, $u^u_k$， $\phi^l_k$, $\phi^u_k$;} %
		\WHILE{$i\leq l$}%
		\STATE{formulate the set $\mathcal{P}_{k,i}$;}%
		\STATE{set $j=1$;}
		\WHILE{$j\leq M_j$}
		\STATE{build two support strips, $\mathcal{S}_{k,i,j}^1$ and $\mathcal{S}_{k,i,j}^2$, overbounding AFSS$_{k,i-1}$ $\cap$ $\mathcal{P}_{k,i}$;}
		\STATE{set $j=j+1$;}
		\ENDWHILE
		\STATE{set $\mathcal{P}'_{k,i}=\bigcap_j(\mathcal{S}_{k,i,j}^1 \cap \mathcal{S}_{k,i,j}^2), j=1,2,...,M_i$;}		
		\STATE{build a zonotope AFSS$_{k,i}$=$\mathcal{Z}_{k,i}^*=p_{k,i}^*\oplus H_{k,i}^*\mathbf{B}^r$ that overbounds AFSS$_{k,i-1}$ $\cap$ $\mathcal{P}'_{k,i}$ with minimal P-radius;}
		\STATE {set i=i+1;}
		\ENDWHILE %
		\STATE {set AFSS$_{k+1,0}$=$\mathcal{Z}_{k,l}^*\oplus\Gamma_{k+1}\mathbf{B}^n=p_{k,l}^*\oplus [H_{k,l}^*~\Gamma_{k+1}]\mathbf{B}^{r+n}$;
		\STATE{set $k=k+1$;}}
		\ENDWHILE
	\end{algorithmic}
\end{algorithm}


Comparing PAZI with CAZI developed in the last section, the major difference lies in the step 13 which constructs candidate zonotopes that overbound the intersection between a zonotope and a polyhedron, that is, AFSS$_{k,i-1}$ $\cap$ $\mathcal{P}'_{k,i}$. In PAZI algorithm, the polyhedron accounts for multiple measurements at once. The details are presented in what follows.

\subsection{Alternative way of building candidate zonotopes}
The parameterization exploited in PAZI algorithm is based on \cite{LE2013}, where a similar bounding approach is applied to a set-membership state estimation problem. This parametrization gives a family of zonotopes which overbound the intersection between a zonotope and a special polyhedron, that is obtained by intersecting a finite number of strips. 

Consider a zonotope $\mathcal{Z}=p\oplus H\mathbf{B}^r\subset \mathbf{R}^n$, a polyhedron given by, 
\begin{eqnarray}
\mathcal{S}'=\{\hat{\theta}\subset \mathbf{R}^n: |\Phi^\top\hat{\theta}-d|\leq \sigma,~\text{given}~ \Phi\in\mathbf{R}^{n\times m}, \nonumber \\ d\in\mathbf{R}^m, \sigma\in \mathbf{R}^m\}.\label{eqn:13_compli}
\end{eqnarray} 
The following parameterization in terms of a matrix $\Lambda\in \mathbf{R}^{n\times m}$ gives a zonotope $\hat{\mathcal{Z}}=\hat{p}\oplus \hat{H}\mathbf{B}^{r+m}\subset \mathbf{R}^n$ such that $\hat{\mathcal{Z}}\supseteq \mathcal{Z}\cap \mathcal{S}'$:
\begin{eqnarray}
\hat{p}(\Lambda) & = & p+\Lambda(d-\Phi^\top p), \nonumber\\
\hat{H}(\Lambda) & = & [(I_n -\Lambda\Phi^\top)H~~\Lambda\Sigma], \label{eqn:13_compli_para}
\end{eqnarray} where $I_n$ is an $n$-dimensional identity matrix and $\Sigma$ is a diagonal matrix given by $\Sigma=diag(\sigma)$.

\begin{figure}[h!]
	\begin{center}
		\includegraphics[width=8cm]{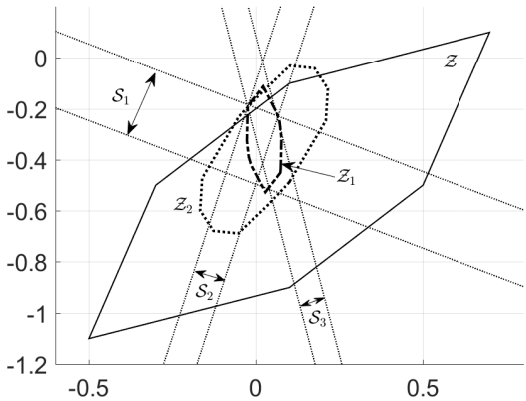}    
		\caption{Example of candidate zonotopes ($\mathcal{Z}_1$ and $\mathcal{Z}_2$).} 
		\label{fig:build_zonos1}
	\end{center}
\end{figure}

As an example, consider a zonotope and a polyhedron defined as follows,
\begin{eqnarray}
\mathcal{Z}=\left[\begin{matrix} \displaystyle
0.1 \\ 
-0.5
\end{matrix}\right]\oplus \left[\begin{matrix} \displaystyle
0.1 & 0.2 & 0.3 \\ 
0.3 & 0.2 & 0.1
\end{matrix} \right]\mathbf{B}^3, \nonumber
\end{eqnarray}
\begin{eqnarray}\displaystyle
\label{eqn:polyhedron}
\mathcal{S'}=\{\hat{\theta}: \left|\left[\begin{matrix} 
5 & 1 \\ 
-4 & 1 \\
1 & 2
\end{matrix} \right]~\hat{\theta}-\left[\begin{matrix} 
-0.1163 \\ 
-0.2935 \\
-0.6928
\end{matrix}\right]\right|\leq\left[\begin{matrix} 
0.2 \\ 
0.2 \\
0.3
\end{matrix}\right] \}.
\end{eqnarray}
Fig.~\ref{fig:build_zonos1} gives an example of two candidate zonotopes, ($\mathcal{Z}_1$ and $\mathcal{Z}_2$), for two different selections of $\Lambda$. Both of them overbound $\mathcal{Z}\cap \mathcal{S'}$, where $\mathcal{S'}=\mathcal{S}_1\cap \mathcal{S}_2\cap \mathcal{S}_3$. But, as shown in Fig.~\ref{fig:build_zonos1}, different choices of $\Lambda$ may give a dramatically different result. The next section discusses how to optimally select $\Lambda$.

\begin{remark}
	In order to apply the above parameterization, the procedure of building support strips is still needed as described in the second step of PAZI algorithm. The polyhedron $\mathcal{S}'$ is obtained by applying \textbf{Proposition 1} $2M_i$ times using all the measurements in the current mini-batch $i$. In other words, $\mathcal{S}'=\mathcal{P}'_{k,i}$ defined in step 12 of PAZI algorithm for each mini-batch. 
\end{remark}

\subsection{Selecting $\Lambda$ based on P-radius}
To assure that the AFSS is not growing unbounded with iterations, certain ``contracting'' properties can be imposed. In this work, we consider the following inequality relation between the results of two neighboring batches, 
\begin{eqnarray}
r_{l}\leq\beta r_{l-1}+\epsilon,\label{eqn:pradius}
\end{eqnarray} 
where for each time step $k$, $r_{l-1}$, $r_{l}$ represent the P-radius of $\mathcal{Z}_{k,l-1}^*$ ($\mathcal{Z}_{k,0}^*=\mathcal{Z}_{k-1,l}^*$) and $\mathcal{Z}_{k,l}^*$, respectively, $\beta\in (0,1]$ is a contraction rate and $\epsilon$ is a positive constant, which, for given $\Gamma'\in \mathbf{R}^{n\times n}$ and $\Sigma'\in \mathbf{R}^{m\times m}$, is determined by the following expression, 
\begin{eqnarray}
\label{eqn:def_epsilon}
\epsilon=\max_{s \in \mathbf{B}^n}{\|\Gamma' s\|_2^2}  + \max_{\eta \in \mathbf{B}^m}{\|\Sigma' \eta\|_2^2}.
\end{eqnarray} 

With $\epsilon$ defined in (\ref{eqn:def_epsilon}), a sufficient condition for (\ref{eqn:pradius}) to hold can be expressed as the following LMI,

\begin{eqnarray}
\label{eqn:LMI}
\mathbf{F}=\left[\begin{array}{cccc}
\beta P & 0 & 0 &  P-\Phi^\top X^\top\\ 
* &\Gamma'^\top \Gamma'  & 0 & \Gamma^\top P-\Gamma^\top\Phi^\top X^\top \\ 
* & * & \Sigma'^\top \Sigma' & \Sigma X^\top \\ 
* & * & * & P 
\end{array} \right]\succeq 0,
\end{eqnarray}
where $X=P\Lambda$, $P$, $\Phi$ and $\Lambda$ are defined in (\ref{eqn:def2}), (\ref{eqn:13_compli}) and (\ref{eqn:13_compli_para}), respectively, $\Gamma$ reflects possible rate of change of parameters between the two iterations and ``$*$" denotes the terms required to ensure the symmetry of $\mathbf{F}$. For the derivation of (\ref{eqn:LMI}), see appendix.

The LMI in (\ref{eqn:LMI}) defines the feasible region for $P$ and $\Lambda$. Specific $P$ and $\Lambda$ are computed by solving the following optimization problem:
\[
\tag{P2}
\begin{array}{rrcl}
\max\limits_{\tau, P, \Lambda} & \tau &          \\
\text{s.t.}~~ {{(1-\beta)P}\over \epsilon} & \succeq & \tau I_n,\\
\mathbf{F} & \succeq & 0,\\
\tau & > & 0.
\end{array}
\]

We solve (P2) using cvx solver \cite{CVX}. 

The P-radius metric was previously exploited for solving state estimation problem in \cite{LE2013}. Our approach here follows the idea from \cite{LE2013} but is adopted to handle mini-batches of measurements, and is applied to parameter estimation where $\Phi$ in (\ref{eqn:LMI}) varies between iterations rather than is a static matrix in the state estimation problem. 

\begin{remark}
	When the number of data points in each batch gets large, the dimension of $~\mathbf{F}$ also increases as well as the complexity of the resulting zonotope. For computational efficiency, we limit $M$ to be less than $10$. At the end of each iteration of PAZI algorithm, the reduction of the order of the zonotope is also needed, which is the same as the one implemented in CAZI algorithm.
\end{remark}

\section{Application to engine condition monitoring}\label{sec:5}
\subsection {Condition monitoring problem}
The proposed algorithms are applied to a marine engine condition monitoring problem. This problem was previously treated in \cite{HW16}, where the compressor and turbine health parameters were simultaneously identified by a specific set-membership identification algorithm. However, in our previous work, overbounding boxes were used to approximate the FSS$_k$, which led to conservative results. 

In order to apply the proposed algorithms, the same model from \cite{HW16} is used here. This model is derived based on the air path dynamics of a marine dual fuel engine. Details of the derivation are found in \cite{HW14}. The dynamics of the turbocharger speed are represented by  
\begin{eqnarray}\label{eq:param2}
\dot{N}_{tc}=\left[\begin{array}{cc}
\displaystyle \theta_t & \displaystyle {1\over\theta_c}
\end{array}\right]
\left[\begin{array}{c}
\displaystyle \phi_1 \\ 
\displaystyle \phi_2
\end{array} \right],
\end{eqnarray} 
where $N_{tc}$ is the turbocharger speed, $\theta_t$ and $\theta_c$ are the two positive constant health parameters to be estimated, $\phi_1$ and $\phi_2$ are given by  
\begin{eqnarray} \label{eq:phi1}
\displaystyle
&\phi_1&={\eta_t^0c_{pe}W_{2t}T_2\over{J_{tc}N_{tc}{\pi\over {30}}}}\left[ 1-\left({p_{amb}\over p_{2}}\right)^{\frac{\gamma-1}{\gamma}}\right],\\
\displaystyle
\label{eq:phi2}
&\phi_2&=-{c_{pa}W_{c1}T_{amb}\over{\eta_c^0J_{tc}N_{tc}{\pi\over {30}}}}\left[\left({p_1\over p_{amb}}\right)^{\frac{\gamma-1}{\gamma}}-1\right],
\end{eqnarray}   
where $\eta_t^0$ and $\eta_c^0$ are nominal turbocharger and compressor efficiencies which vary in time and are known, $c_{pa}$ and $c_{pe}$ are the specific heats at constant pressure of air and exhaust gas, respectively, $W_{2t}$ and $W_{c1}$ are the mass flow rates through the compressor and through the turbine, respectively, $T_2$ and $T_{amb}$ are the exhaust manifold temperature and the ambient temperature, respectively, $p_1$, $p_2$ and $p_{amb}$ are the intake manifold pressure, the exhaust manifold pressure and the ambient pressure, respectively, $\gamma$ is the ratio of specific heats, and $J_{tc}$ is the turbocharger inertia.

Since the flow measurements are not available in marine engines, we treat $W_{c1}$ and $W_{2t}$ as signals with unknown but bounded uncertainties, whose upper and lower bounds may be estimated \cite{J.Bravo2006}, \cite{HW16}, i.e.,
\begin{eqnarray}
\label{eqn:Wc1}
&0&< W_{c1}^l(t)\leq W_{c1}(t)\leq W_{c1}^u(t),\\
\label{eqn:W2t}
&0&< W_{2t}^l(t)\leq W_{2t}(t)\leq W_{2t}^u(t).
\end{eqnarray}

As we can see from (\ref{eq:param2}), (\ref{eq:phi1}) and (\ref{eq:phi2}), the flow uncertainties affect the regressors and are, therefore, multiplicative. By assuming that the rest of the variables, except for $\theta_t$ and $\theta_c$, which we treat as unknown constant parameters, is known either through measurements or accurate estimation, the bounds on the regressors can be obtained by 
\begin{eqnarray}
\label{bnd:phi1}
\phi_1^l=\phi_1(W_{2t}^l),~\phi_1^u=\phi_1(W_{2t}^u),\\
\label{bnd:phi2}
\phi_2^l=\phi_1(W_{c1}^u),~\phi_2^u=\phi_1(W_{c1}^l).
\end{eqnarray}
The measurement noise associated with the turbocharger speed measurement is handled by an input observer, which provides the estimate of the time rate of change of $N_{tc}$ and the error bounds. Then, in order to match the general problem setting in (\ref{eqn:1}), we set
\begin{eqnarray}
\label{eqn:21}
y_k &=& \hat{\dot{N}}_{tc}(k),\nonumber\\
\label{eqn:22}
\theta_k &=&[\theta_t~ \displaystyle {1\over\theta_c}]^\top,\nonumber\\
\label{eqn:23}
\phi_k&=&[\phi_1(k)~\phi_2(k)]^\top,\nonumber\\
\label{eqn:24}
u_k&=&e(k)=\dot{N}_{tc}(k)-\hat{\dot{N}}_{tc}(k),
\end{eqnarray}
where $\hat{\dot{N}}_{tc}(k)$, $e(k)$ represent the estimated time rate of change of $N_{tc}$ from the input observer and its estimation error. Details of the design of the input observer and the computation of the error bounds may be found in  \cite{IK06}, \cite{HW16}.

In practice, the health parameters vary slowly as the performance of the components degrade during engine's life time due to aging. The identification of the health parameters is performed at different time instances during its life span by collecting corresponding measurements, which are indicative of current health condition of the engine. For our application, we treat the health parameters as constant for each batch of measurements collected in a given time period. It is assumed that the bounds on the rates of parameter change, i.e., $\Gamma$, are known. Note that each batch of measurements could be further divided into smaller batches in order to reduce the computational complexity as described in \textbf{Remark 4}.    

\subsection{Results and discussions}
\begin{figure}[h!]
	\begin{center}
		\includegraphics[width=9cm]{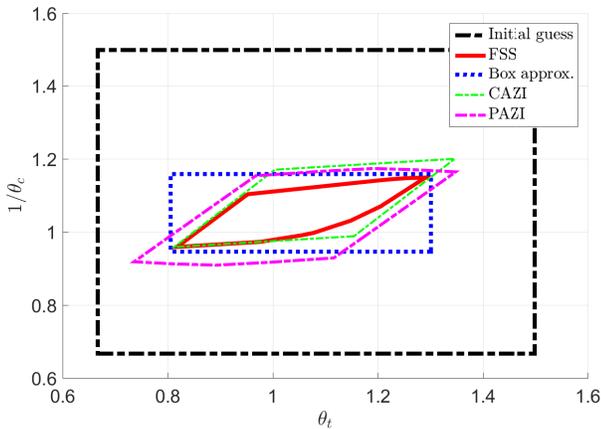}    
		\caption{Comparison of the estimation results when using box overbounding \cite{HW16} and the proposed algorithms which use zonotopes.} 
		\label{fig:app_results1}
	\end{center}
\end{figure}

For the same data set as used in \cite{HW16}, which consists of 1500 data points, Fig.~\ref{fig:app_results1} shows the estimation results after the final iteration. For the PAZI algorithm, the number of data points in each mini-batch is limited to $4$, that is, $M_i=4$, $i=1,2,...,375$. The true parameters remains constant, that is, $\theta_k=[1~1]^\top$. Note that both the CAZI algorithm and the PAZI algorithm provide a tight overbound of the FSS, where the latter is computed by a general constraint elimination function \cite{MPT}. As compared with  \cite{HW16} that used boxes to approximate the FSS$_k$, the algorithms developed in this paper provide more accurate estimation performance if the results are judged based on the volume of AFSS. The volumes of different approximations as well as of the FSS can be found in Table~\ref{tb:volcomp}. The computation times based on a $2.2$ GHz Windows machine are also reported in Table~\ref{tb:volcomp}.

\begin{remark}
	Infeasiblility of the LMIs condition (\ref{eqn:LMI}) has been observed for the PAZI algorithm, when the condition number of $\Phi$ or the trace of $(\Sigma'\Sigma)$ gets large enough. Note that $cond(\Phi)$ and $tr(\Sigma'\Sigma)$ are related to the signal richness and the uncertainty level, respectively. If infeasibility is encountered, the PAZI algorithm only applies the prior estimate to AFSS.
\end{remark}

\begin{figure}[h!]
	\begin{center}
		\includegraphics[width=9cm]{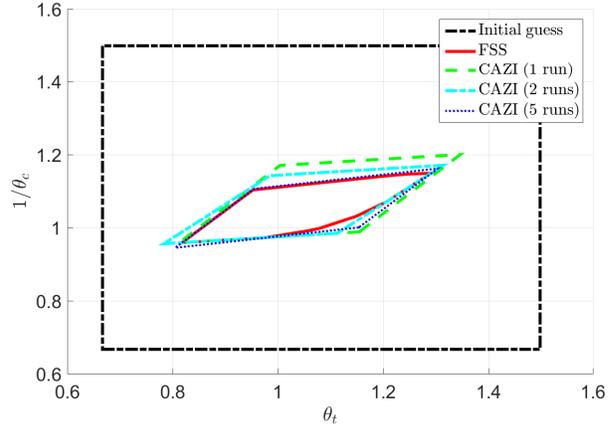}    
		\caption{Improvement of the estimation accuracy by running CAZI multiple times over the same data set.} 
		\label{fig:app_results2}
	\end{center}
\end{figure}

\begin{table}[h!]
	\centering
	\caption{Volumes and CPU times of the FSS, AFSS based on boxes \cite{HW16} and AFSSs for the zonotope-based solutions (CAZI and PAZI) proposed in this paper.} 
	\label{tb:volcomp}
	\begin{tabular}{|c|c|c|c|c|c|}
		\hline
		\textbf{Case}   & FSS    & Box approx.    & CAZI & PAZI \\ \hline
		\textbf{Volume} & 0.0414 & 0.1053 & 0.0668  & 0.0929\\ \hline
		\textbf{\begin{tabular}{@{}c@{}}CPU time \\ (sec)\end{tabular}} & 145 & 39 & 47  & 98\\ \hline
	\end{tabular}
\end{table} 
Furthermore, the estimation accuracy may be further improved by running the proposed algorithms multiple times through the same data set (while using final estimate of the previous run as an initial estimate for the next run). Fig.~\ref{fig:app_results2} illustrates an example of applying the CAZI algorithm multiple times. Doing so two times reduces the volume of AFSS to 0.0556. After five times, the volume is reduced to 0.0479 which is a 28\% reduction compared to applying the algorithm just once. No further significant reduction appears possible with more than five repetitions, nor a similar procedure is able to improve the accuracy of the box-based AFSS. 

\section{Conclusions}\label{sec:6}
In this paper, two set-membership estimation algorithms (CAZI and PAZI) for the identification of time-varying parameters in linear models have been developed, which account for both additive and multiplicative uncertainties. The two algorithms can be discriminated by their ways of processing the data. CAZI processes one measurement at a time while PAZI processes multiple measurements at each iteration. Their computational loads are also different. CAZI involves solving linear programming problems while PAZI requires solving linear programming problems as well as an LMI problem for each iteration with the help of a convex optimization solver. However, PAZI offers an opportunity of developing bounds on the P-radius of AFSS produced by the algorithm in each iteration. We have demonstrated that both algorithms have the capability to provide tight approximated feasible solution sets (AFSSs), which are zonotopes. An application of the algorithms to an engine condition monitoring problem was reported to demonstrate the use of the proposed algorithms. 

\begin{ack}                               
The authors would like to thank Prof. Diego Regruto from Politecnico di Torino for valuable discussions and comments on this work.  
\end{ack}

\bibliographystyle{plain}        
\bibliography{autosam}           

\begin{thebibliography}{10}

\bibitem{J.Bravo2006}
J.~M. Bravo, T.~Alamo, and E.~F. Camacho.
\newblock Bounded error identification of systems with time-varying parameters.
\newblock {\em IEEE Transactions on Automatic Control}, 51:1144--1150, 2006.

\bibitem{Cas14}
M.~Casini, A.~Garulli, and A.~Vicino.
\newblock Feasible parameter set approximation for linear models with bounded
  uncertain regressors.
\newblock {\em IEEE Transactions on Automatic Control}, 59:2910--2920, 2014.

\bibitem{Piga12}
V.~Cerone, D.~Piga, and D.~Regruto.
\newblock Set-membership error-in-variables identification through convex
  relaxation techniques.
\newblock {\em IEEE Transactions on Automatic Control}, 57:517--522, 2012.

\bibitem{Ble16}
R.~M. Fernández-Cantí, V.~Puig J.~Blesa, and S.~Tornil-Sin.
\newblock Set-membership identification and fault detection using a bayesian
  framework.
\newblock {\em Int. J. Adapt. Systems Science}, 47:1710--1724, 2016.

\bibitem{Puig15}
R.~M. Fernández-Cantí, S.~Tornil-Sin, J.~Blesa, and V.~Puig.
\newblock Non-linear set-membership identification approach based on the
  bayesian framework.
\newblock {\em IET Control Theory \& Applications}, 9:1392--1398, 2015.

\bibitem{Cas16}
A.~Garulli and A.~Vicino.
\newblock A linear programming approach to online set membership parameter
  estimation for linear regression models.
\newblock {\em Int. J. Adapt. Control Signal Process.}, 31:360--378, 2016.

\bibitem{CVX}
Michael Grant and Stephen Boyd.
\newblock Cvx: Matlab software for disciplined convex programming, version 2.1.
\newblock \url{http://cvxr.com/cvx}, March 2014.

\bibitem{Guo15}
Y.~Guo, Y.~Zhang, and B.~Jiang.
\newblock Set-membership estimation-based adaptive reconfiguration scheme for
  linear systems with disturbances.
\newblock {\em Int. J. Adapt. Control Signal Process.}, 30:359--374, 2016.

\bibitem{MPT}
M.~Herceg, M.~Kvasnica, C.N. Jones, and M.~Morari.
\newblock {Multi-Parametric Toolbox 3.0}.
\newblock In {\em Proc.~of the European Control Conference}, pages 502--510,
  Z\"urich, Switzerland, July 17--19 2013.
\newblock \url{http://control.ee.ethz.ch/~mpt}.

\bibitem{Her15}
P.~Herrero, B.~Delaunay, L.~Jaulin, N.~Oliver P.~Georgiou, and C.~Toumazou.
\newblock Robust set-membership parameter estimation of the glucose minimal
  model.
\newblock {\em Int. J. Adapt. Control Signal Process.}, 30:173--185, 2016.

\bibitem{IK06}
I.~Kolmanovsky, I.~Sivergina, and J.~Sun.
\newblock Simultaneous input and parameter estimation with input observers and
  set-membership parameter bounding: theory and an automotive application.
\newblock {\em Int. J. Adapt. Control Signal Process.}, 20:225--246, 2006.

\bibitem{KUR97}
A.~Kurzhanski and I.~Valyi.
\newblock {\em Ellipsoidal Calculus for Estimation and Control}.
\newblock Springer, 1997.

\bibitem{LE2013}
V.~T.~H. Le, C.~Stoica, T.~Alamo, E.~F. Camacho, and D.~Dumur.
\newblock {\em Zonotopes: From Guaranteed State-estimation to Control}.
\newblock John Wiley \& Sons, 2013.

\bibitem{LB69}
D.G. Luenberger.
\newblock {\em Optimization by vector space methods}.
\newblock John Wiley \& Sons, 1969.

\bibitem{Milanese1996}
M.~Milanese and A.~Vicino.
\newblock Optimal estimation theory for dynamic systems with set membership
  uncertainty: an overview.
\newblock In M.~Milanese, J.~Norton, H.~Piet-Lahanier, and E.~Walter, editors,
  {\em Bounding Approaches to System Identification}, pages 5--27. Springer,
  1996.

\bibitem{Ping15}
X.~Ping and N.~Sun.
\newblock Dynamic output feedback robust model predictive control via zonotopic
  set-membership estimation for constrained quas-lpv systems.
\newblock {\em Int. J. of Applied Mathematics}, 2015.

\bibitem{Rot14}
D.~Rotondo, F.~Nejjari, V.~Puig, and J.~Blesa.
\newblock Model reference ftc for lpv systems using virtual actuators and
  set-membership fault estimation.
\newblock {\em Int. J. Robust and Nonlinear Control}, 25:735--760, 2015.

\bibitem{Yu15}
E.~Zamora W.~Yu and A.~Soria.
\newblock Ellipsoid slam: a novel set membership method for simultaneous
  localization and mapping.
\newblock {\em Autonomous Robots}, 40:125--137, 2016.

\bibitem{HW14}
H.~Wang, I.~Kolmanovsky, and J.~Sun.
\newblock Feedback control during mode transition for a marine dual fuel
  engine.
\newblock {\em IFAC-PapersOnLine}, 48:279--284, 2015.

\bibitem{HW16}
H.~Wang, I.~Kolmanovsky, and J.~Sun.
\newblock Set-membership condition monitoring framework for dual fuel engines.
\newblock In {\em 2016 American Control Conference (ACC)}, pages 3298--3303,
  July 2016.

\bibitem{HW17}
H.~Wang, I.~Kolmanovsky, and J.~Sun.
\newblock Set-membership parameter identification of linear systems with
  multiplicative uncertainties: A new algorithm.
\newblock In {\em 2017 American Control Conference (ACC) (to appear)}, May
  2017.

\bibitem{Wei15}
G.~Wei, S.~Liu, Y.~Song, and Y.Liu.
\newblock Probability-guaranteed set-membership filtering for systems with
  incomplete measurements.
\newblock {\em Automatica}, 60:12--16, 2015.

\end{thebibliography}



\appendix
\section{Appendix}\label{sec:7}
\subsection{Proof of Proposition 1}
\begin{pf} 
	The proof of Proposition 1 can be completed in two steps. 
	The first step is to prove that with $\delta=\delta_1^*$, the strip $\mathcal{S}^1$ is a superset of $\mathcal{P}\cap \mathcal{C}$. This is equivalent to showing that for any  $\hat{\theta} \in \mathcal{P}\cap \mathcal{C}$, $\hat{\theta} \in \mathcal{S}^1$. Clearly, $\hat{\theta}\in \mathcal{P}\cap \mathcal{C}$ is a feasible solution to the optimization problem (P1) and $|{\tfrac{(\hat{\theta}-\tilde{\theta}_1)^\top\phi_1}{\|\phi_1\|}}|\leq \delta_1^*$. Given that $\phi^\top_1\tilde{\theta}_1= b_1$, we can rewrite the definition of $\mathcal{S}^1$ in (\ref{eqn:14}) with $\delta=\delta_1^*$ as  
	\begin{eqnarray}
	\label{sk1_2}
	\mathcal{S}^1=\{\hat{\theta}~|~0\leq (\hat{\theta}-\tilde{\theta}_1)^\top\phi_1\leq \delta_1^* \|\phi_1\|\}.
	\end{eqnarray}
	
	Note that $(\hat{\theta}-\tilde{\theta}_1)^\top\phi_1\geq 0$ because $\hat{\theta} \in \mathcal{C}$. Furthermore, $(\hat{\theta}-\tilde{\theta}_1)^\top\phi_1\leq\delta_1^* \|\phi_1\|$ based on (\ref{sk1_2}). Thus we conclude that $\hat{\theta} \in \mathcal{S}^1$, when $\delta=\delta_1^*$.
	
	The second step is to prove that $\delta=\delta_1^*$ is the smallest value of $\delta$ for $\mathcal{S}^1$ to be a superset of $\mathcal{P} \cap \mathcal{C}$. The proof is by contradiction. Assume there exists $\epsilon>0$ such that $\delta=\delta_1^*-\epsilon$ can also be used to define such a strip $\mathcal{S}^1$ with the following representation,
	\begin{eqnarray}
	\label{sk1_3}
	\mathcal{S}^1=\{\hat{\theta}~|~0\leq (\hat{\theta}-\tilde{\theta}_1)^\top\phi_1\leq (\delta_1^*-\epsilon) \|\phi_1\|\}.
	\end{eqnarray}
	
	Consider
	\begin{eqnarray}
	\hat{\theta}_{max}\in \arg\max_{\hat{\theta}     \in  \mathcal{P}\cap \mathcal{C}}|{\tfrac{(\hat{\theta}-\tilde{\theta}_1)^\top\phi_1}{\|\phi_1\|}}|.
	\end{eqnarray}
	
	Clearly, $\hat{\theta}_{max} \in  \mathcal{P}\cap \mathcal{C}$. By the definition of the optimization problem (P1), the cost associated with $\hat{\theta}_{max}$ is $\delta_1^*=|  {\tfrac{(\hat{\theta}_{max}-\tilde{\theta}_1)^\top\phi_1}{\|\phi_1\|}}|$. Then $(\hat{\theta}_{max}-\tilde{\theta}_1)^\top\phi_1= \delta_1^* \|\phi_1\|> (\delta_1^*-\epsilon) \|\phi_1\|$ and $\hat{\theta}_{max}\notin \mathcal{S}^1$. Thus $\mathcal{S}^1$ in (\ref{sk1_3}) is not a superset of $\mathcal{P}\cap \mathcal{C}$.
	
\end{pf}

\subsection{Derivation of the LMI}
The LMI condition in (\ref{eqn:LMI}), $\mathbf{F}\succeq0$, is derived as follows. 

Apply the definition in (\ref{eqn:def2}) to (\ref{eqn:pradius}) and substitute the $\epsilon$ determined by (\ref{eqn:def_epsilon}), we get,
\begin{eqnarray}
\label{eqn:LMIproof1}
\max_{z \in \mathbf{B}^{r+n+m}}~\|H_l^* (\Lambda)z\|_P^2  \leq  \beta \max_{z' \in \mathbf{B}^r}~\|H_{l-1}^* z'\|_P^2 \nonumber \\  + \max_{s \in \mathbf{B}^n}{\|\Gamma' s\|_2^2}  + \max_{\eta \in \mathbf{B}^m}{\|\Sigma' \eta\|_2^2}.
\end{eqnarray} 

An sufficient condition for (\ref{eqn:LMIproof1}) to hold is:
\begin{eqnarray}
\label{eqn:LMIproof2}
\max_{z \in \mathbf{B}^{r+n+m}}~(\|H_l^* (\Lambda)z\|_P^2 - \beta~\|H_{l-1}^* z'\|_P^2 \nonumber \\
-{\|\Gamma' s\|_2^2}-{\|\Sigma' \eta\|_2^2}) \leq 0,
\end{eqnarray} 
with $z=[z'^\top s^\top \eta^\top]^\top \in \mathbf{B}^{r+n+m}$, $\eta \in \mathbf{B}^m$.

It then follows that, (\ref{eqn:LMIproof2}) is equivalent to the following inequality:
\begin{eqnarray}
\label{eqn:LMIproof3}
z^\top H_l^{* \top}(\Lambda)PH_l^* (\Lambda)z - \beta z'^\top H_{l-1}^{* \top}PH_{l-1}^* z' \nonumber \\ -s^\top\Gamma'^\top  \Gamma' s-\eta^\top\Sigma'^\top \Sigma' \eta \leq 0,~~\forall z', s, \eta.
\end{eqnarray}  

The explicit form of $H_l^* (\Lambda)$ is then given by applying the parameterization in (\ref{eqn:13_compli_para}) with $H=[H_{l-1}^*~\Gamma]$, 
\begin{eqnarray}
\label{eqn:LMIproof4}
H_l^{*}(\Lambda)=[(I_n -\Lambda\Phi^\top)H_{l-1}~~(I_n -\Lambda\Phi^\top)\Gamma~~\Lambda\Sigma].
\end{eqnarray}

Right multiplying by $z=[z'^T s^T \eta^T]^T$ yields
\begin{eqnarray}
\label{eqn:LMIproof5}
H_l^* (\Lambda)z = ((I_n -\Lambda\Phi^\top))(H_{l-1}^*z'+\Gamma s)+ \Lambda\Sigma\eta.
\end{eqnarray} 

Denote $\bar{z}=H_{k-1}^*z'$ and substitute the expression from (\ref{eqn:LMIproof5}) into (\ref{eqn:LMIproof3}). Then the following matrix inequality can be obtained, which is equivalent to (\ref{eqn:LMIproof3}):
\begin{eqnarray}
\label{eqn:LMIproof7}
\left[\begin{matrix}
\bar{z} \\ 
s \\ 
\eta
\end{matrix}\right]^T
\left(A - B P^{-1} B^\top \right)\left[\begin{matrix}
\bar{z} \\ 
s \\ 
\eta
\end{matrix}\right] \geq 0, \forall \bar{z}, s, \eta,
\end{eqnarray}
where
\begin{eqnarray*}
	A=\left[\begin{matrix}
		\beta P & 0 & 0 \\ 
		* & \Gamma'^\top \Gamma' & 0 \\ 
		* & * & \Sigma'^\top \Sigma' 
	\end{matrix}\right], B= \left[\begin{matrix}
		(I-\Phi^\top\Lambda^\top)P \\ 
		(\Gamma^\top-\Gamma^\top\Phi^\top\Lambda^\top)P \\ 
		\Sigma\Lambda^\top P
	\end{matrix}\right].
\end{eqnarray*}

By applying Schur complement to (\ref{eqn:LMIproof7}), it follows that if (\ref{eqn:LMI}) holds then (\ref{eqn:LMIproof1}) and (\ref{eqn:pradius}) hold. 
\end{document}